# SRFS: Parallel Processing Fault-tolerant ROS2-based Flight Software for the Space Ranger CubeSat

Zebei Zhao, Yinghao Xiang, Ziyu Zhou, Kehan Chong, Haoran Ma, Pei Chen

*Abstract*—**Traditional Real-Time Operating Systems (RTOS) often suffer from limited parallel performance, whereas thread monitoring in Linux-based systems remains challenging. To overcome these limitations, this paper presents a satellite flight software system design based on the Robot Operating System (ROS), which utilizes its reliable built-in publish-subscribe messaging mechanism to facilitate inter-application communication. In response to the complex functional demands of modern small satellites, the proposed design integrates both hardware and software architectures, along with system scheduling and error-correction strategies. This integration supports efficient parallel data processing, enhances system reliability, and shortens the development cycle through code reuse. The system was rigorously evaluated through comprehensive tests covering time delay, system management, fault tolerance, and maintenance procedures. Experimental results confirm the system's effectiveness in telemetry, remote control, integration of new features, and autonomous error recovery. The findings underscore the high reliability and maintainability of the ROS-based satellite flight software, offering a valuable reference for the rapid development of high-performance small satellite systems.**

## I. INTRODUCTION

With the increasing frequency of low-orbit space exploration missions, low-cost small satellites are being widely adopted. At the same time, small satellite missions are becoming more complex, requiring the processing of growing volumes of data. Therefore, onboard data processing has become a trend in the development of small satellites[1]. For image data, complex algorithms are required to extract useful information, particularly those used in AI models, which demand substantial computational power [2]. This places higher requirements on the computational capacity of the onboard processors and the task scheduling capabilities of the satellite flight software.

Existing small satellites cover a wide range of onboard computer processors, including x86 [3], ARM [4], SPARC [5], PowerPC [6], Loongson [7], Cambrian [8], graphics processors (GPU) [9], and high-performance Field-Programmable Gate Arrays (FPGA) [10]. For low-cost small satellites, ARM processors are widely used, with applications ranging from microprocessors such as STM32 [11] to high-performance processors like the Nvidia Tegra X2/X2i [12]. Regarding operating systems, small satellites generally use RTOS, such as VxWorks [13], μC/OS-II [14]and FreeRTOS [15]. However, with the increasing computational demands of processors, the once commonly used Microcontroller Units (MCUs) no longer have sufficient computational power to handle tasks such as image processing. The RTOS associated with these MCUs also struggle to schedule high-computational tasks, like image processing, while simultaneously handling regular satellite telemetry and telecommand tasks. Therefore, more powerful multi-core embedded processors are required.

For multi-core embedded processors, developing the spacecraft software based on the Linux operating system is a feasible approach [16]. The Aalto-1 satellite, which is a nanosatellite designed and built by students and researchers at Aalto University, has adopted this solution [17]. However, Linux is not inherently designed for satellite software scheduling and satellite communication. Designing satellite operations software based on Linux requires a significant amount of effort to establish inter-software communication and address reliability issues caused by Linux system processes being terminated unexpectedly.

Therefore, for multi-core processors using a Linux system, adopting a flight software designed for inter-process and hardware communication is a better choice. For satellite flight software suitable for multi-core Linux systems, NASA's F Prime, which supports multi-core processors, offers a feasible option. However, the multi-core scheduling of the F Prime system remains unverified, leaving its reliability uncertain [18].

This work was supported in part by the Fundamental Research Funds for the Central Universities of China under Grant 501XSKC2024115001 and the Beijing Natural Science Foundation under Grant QY25227. *(Corresponding author: Pei Chen).*

Zebei Zhao was with the School of Automation Science and Electrical Engineering, Beihang University, Beijing 100191, China. He is now with the School of Artificial Intelligence and Data Science, University of Science and Technology of China, Hefei 230026, China (e-mail: zbzhao@mail.ustc.edu.cn).

Yinghao Xiang is with the School of Computer Science and Engineering, Beihang University, Beijing 100191, China (e-mail: xiangyinghao@buaa.edu.cn).

Ziyu Zhou is with the School of Astronautics, Beihang University, Beijing 100191, China, and also with the Key Laboratory of Spacecraft Design Optimization and Dynamic Simulation Technologies (Ministry of Education), China (e-mail: zzy010710@buaa.edu.cn).

Kehan Chong is with the School of Astronautics, Beihang University, Beijing 100191, China, and also with the Key Laboratory of Spacecraft Design Optimization and Dynamic Simulation Technologies (Ministry of Education), China (e-mail: chongkehan@buaa.edu.cn).

Haoran Ma is with the School of Astronautics, Beihang University, Beijing 100191, China (e-mail: mahaoran_bh@buaa.edu.cn).

Pei Chen is with the School of Astronautics, Beihang University, Beijing 100191, China, and also with the Key Laboratory of Spacecraft Design Optimization and Dynamic Simulation Technologies (Ministry of Education), China (e-mail: chenpei@buaa.edu.cn).

The Robot Operating System (ROS), an open-source operating system for robots developed by Willow Garage, was designed to enhance code reusability and modularity in increasingly complex robotics applications [19]. Based on the current applications of ROS in aerospace, [20] presents one of the initial attempts to use ROS as a flight software framework on nanosatellites. This application has been demonstrated in the context of two CubeSat platforms: the Drag De-Orbit Device (D3) and the Passive Thermal Coating Observatory Operating in Low Earth Orbit (PATCOOL) [21]. In [22], the Japan Aerospace Exploration Agency (JAXA) combines ROS with NASA's Core Flight System (cFS) to establish an open-source framework intended to directly port robotics applications from ground-based robots to spacecraft. However, the communication between nodes in ROS1 relies on a unique ROS master node for management. Consequently, if this node fails, the entire ROS system becomes inoperable. This limitation makes it challenging for ROS1 to be applied in environments with stringent reliability requirements. Recognizing this issue, the ROS development team introduced the second-generation Robot Operating System (ROS2) to address these concerns.

ROS2 is based on the open Data Distribution Service (DDS) communication standard, offering top-notch security and strong real-time capabilities, as well as support for embedded devices [23]. According to the official ROS design documentation, the DDS middleware used in ROS2 features end-to-end communication, which allows any two DDS programs to communicate directly without the need for tools like the ROS Master. This design enhances the system's fault tolerance and flexibility[24].

For ROS2 applications in space, [25] describes an attempt by Politecnico di Torino to develop a small satellite flight control system using a Raspberry Pi with ROS2. However, this work only involves basic software development for tasks such as bus data reading and attitude control and does not incorporate a fully developed scheduling or error correction mechanism. In addition, [26] describes the implementation and performance of the Micro-ROS/ROS2 framework in the design of attitude determination and control system algorithms. However, the research is limited to the implementation of the attitude control part and does not extend to the entire satellite software system. The open-source framework Space Station OS, developed by the Japanese company SpaceData, attempts to apply the ROS2 system to space stations. However, it currently only provides a macroscopic simulation of the various subsystems of the space station, lacking detailed implementation solutions[28].

Furthermore, [27] discusses NASA's development of Space ROS, which attempts to bridge ROS2 with its original F prime software system in order to leverage ROS2 for controlling applications like the Mars rover and space robotic arm. However, this bridging method discards ROS2's most important system-level feature: its decentralized architecture, which lacks a special master node, thus preventing a single point of failure and granting ROS2 natural reliability. Simultaneously, this integration approach introduces additional latency overhead at the interface layer, and the system's reliability becomes entirely dependent on the stability of the bridge's communication. Once this communication fails, the large number of payloads controlled by ROS2 faces a significant risk of failure.

In summary, current literature lacks a mature and reliable ROS2 flight software for spacecraft operations systems, and no studies to date have demonstrated practical engineering implementation or operational readiness.

The Space Ranger Satellite (SR-SAT), a CubeSat designed for space debris surveillance, is equipped with a smart camera and an impact-based payload[29]. Notably, it is China's first CubeSat specifically developed for space debris surveillance. The space debris detection mission requires SR-SAT to perform onboard image processing in parallel with routine satellite management, which has driven the design of its software and electronic systems[30]. Our Processing Fault-tolerant ROS2-based Flight Software is specifically designed for SR-SAT and is accordingly named the Space Ranger Flight Software (SRFS). To the best of our knowledge, SRFS is the first satellite flight software in China developed using ROS2. We have open-sourced part of the source code for our software architecture[31].

The primary contributions of this paper are as follows:

(1) a scheduling mechanism tailored for complex small satellite missions involving image processing, enabling parallel processing of short-duration, high-computation tasks alongside long-duration, strictly time-sequenced tasks

(2) a ROS2-based fault-tolerance mechanism using information flow monitoring, providing autonomous error correction across multiple scenarios

(3) a software maintenance mechanism that supports modification and addition of software programs while the system is operational.

This paper will elaborate on the hardware and software architecture of a ROS-based satellite flight software for small satellites, introducing a framework on single processor chip (with 4 cores) utilizing ROS nodes. It details the system scheduling process and presents solutions for handling unforeseen errors, thereby demonstrating the reliability and advantages of the proposed approach. Additionally, we simulate node operations to develop and validate a complete ROS2-based satellite flight software specifically suited for small satellites, including testing and verification to confirm system effectiveness.

## II. Overview of the Parallel Processing Fault-Tolerant Flight Software Tasks

The primary tasks of the parallel satellite flight software include satellite management, telemetry and telecommand (TT&C), and image acquisition and processing. As shown in Figure 1, the three main tasks undergo initialization after system startup. Upon completing initialization, the TT&C and satellite management tasks run in parallel, while the image acquisition and processing task is triggered and executed upon receiving a TT&C command.

The primary functions of the TT&C operations include the transmission of telemetry data and the execution of telecommand tasks. Upon receiving a telemetry downlink command, the satellite management system retrieves stored payload operation data for a specific period, encapsulates this information into data frames, and sends it to the TT&C subsystem to complete the telemetry downlink task. Upon receiving telecommand instructions, the satellite management system configures and executes corresponding tasks as directed.

The image acquisition and processing tasks consist of three main components: image acquisition, image storage, and image processing. Upon receiving an image acquisition command through a remote signal, the satellite management system sets up an image acquisition task, initiating the onboard CMOS sensor to capture images at the designated time and store them in the appropriate storage space. Subsequently, the image processing node reads and processes the stored images. Of the three components involved in this task, a RTOS based on an MCU can handle image acquisition and storage, but it is insufficient for image processing.

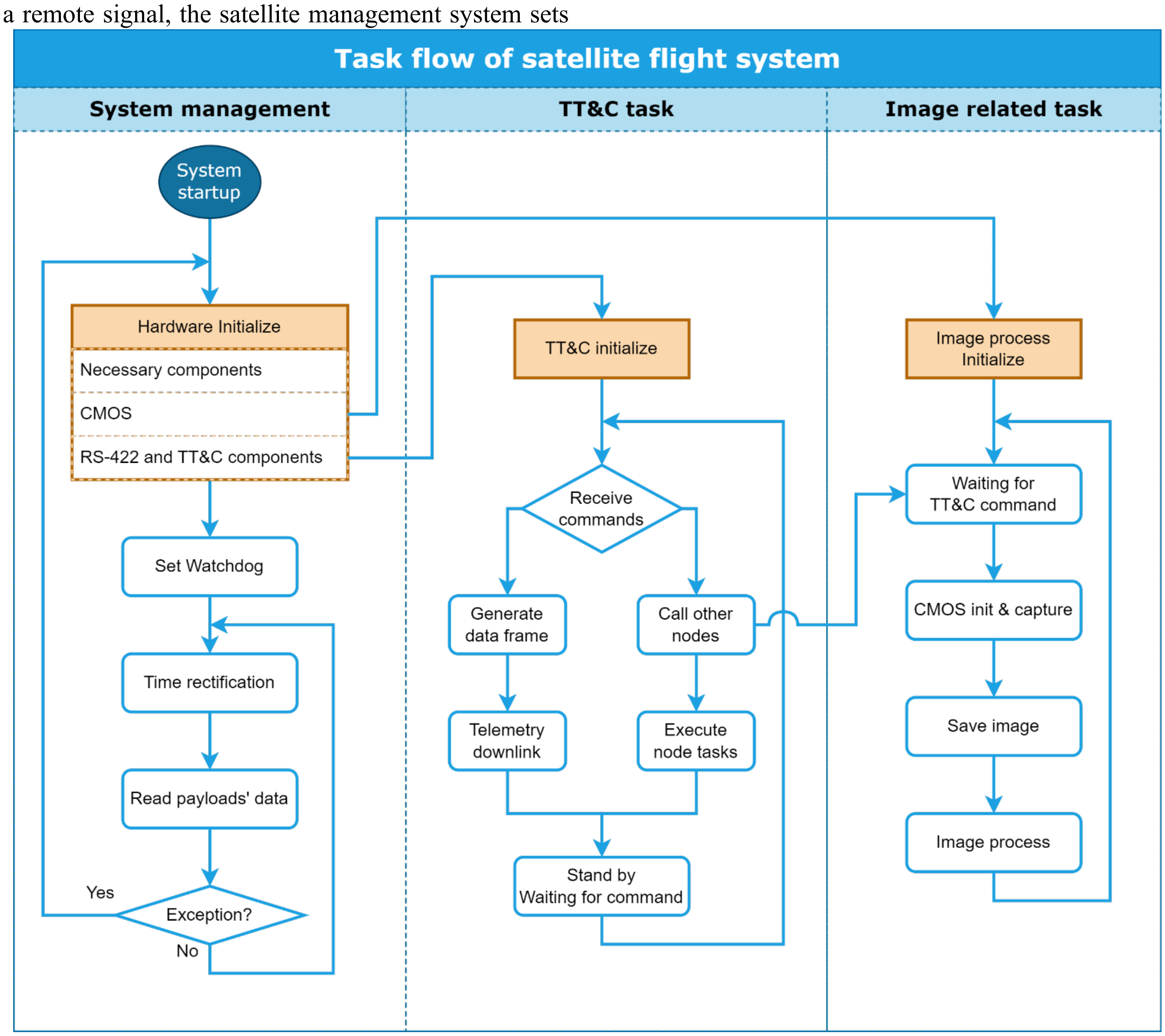

**Fig. 1.** Task flow of satellite flight system

Therefore, a high-performance processor running a Linux system is required to handle image processing tasks. Leveraging existing image processing algorithms within the ROS ecosystem on Linux enables rapid deployment of image processing tasks, thereby enhancing development efficiency.

In summary, to address the challenges posed by parallel processing of high-volume image acquisition and processing tasks alongside time-sensitive satellite management operations, we integrated essential satellite management tasks with the TT&C, image acquisition, and processing payload tasks into a single multi-core processor. These were organized as separate task nodes within the ROS system, with inter-task collaboration logic designed to enable multitasking concurrency and scheduling within a single multi-core processor. This approach has allowed us to build a satellite management system with multi-task parallel processing capabilities.

## III. Implementation Scheme for the Parallel Processing Fault-Tolerance Flight Software

As the sole information processing system for a small satellite, the parallel fault-tolerant flight software must frequently interact with external hardware, retrieving and transmitting hardware data. The software's functionality is designed based on the existing hardware configuration; therefore, the following implementation scheme outlines both the hardware architecture and software architecture.

### *A. Hardware Architecture*

Traditional MCUs and FPGAs cannot meet the requirements for high data volumes and parallel-thread tasks, necessitating the selection of a more powerful processor. Rockchip's RK series processors, currently among China’s leading IoT and AIoT processor chips, fulfill the performance demands of image processing. Given the power and size constraints of CubeSats, the parallel, fault-tolerant flight software adopts the RK3568 processor as the onboard computer. This processor balances high computational capacity for image processing with low power consumption requirements [31].

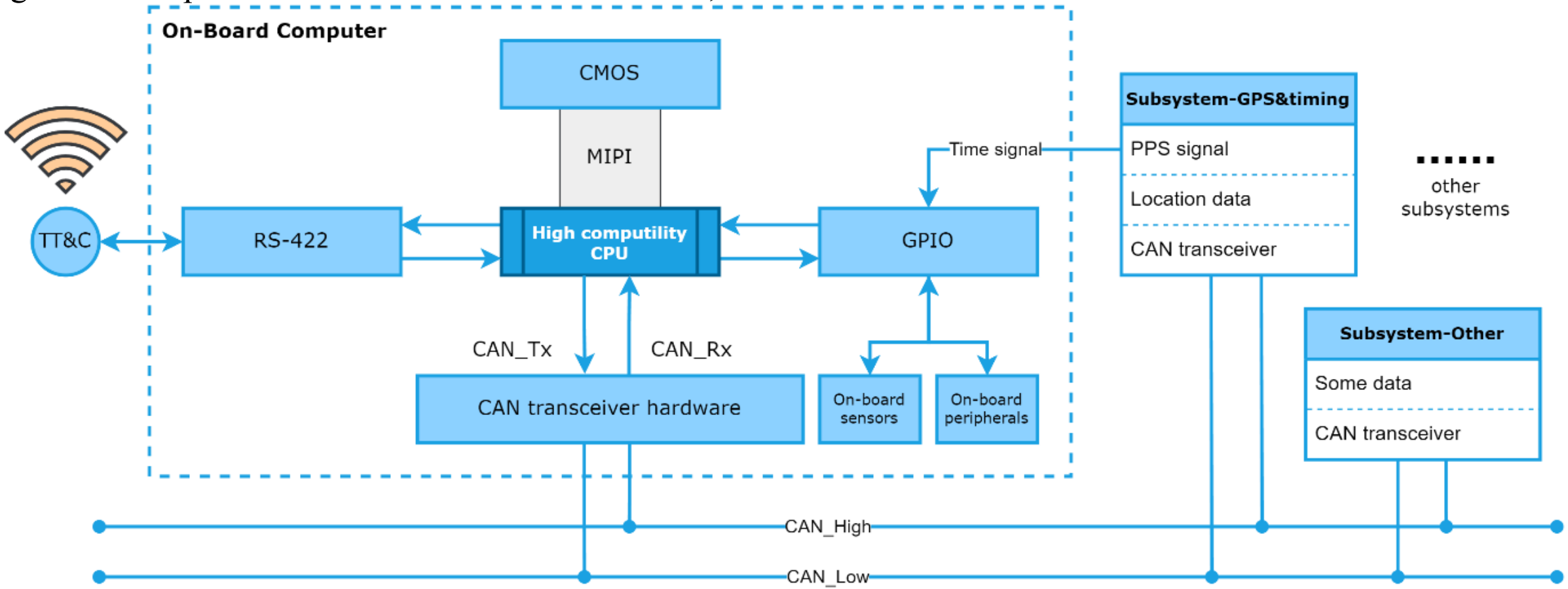

**Fig. 2.** Hardware architecture of satellite housekeeping system

At the hardware connectivity level, the onboard computer communicates with the payload through a CAN bus, performing time-division multiplexing to handle data exchange with various peripherals. It connects to the onboard CMOS sensor via an MIPI bus and interfaces with the TT&C subsystem through an RS422 interface, enabling downlink of telemetry data and uplink of telecommand signals. Additionally, signals from onboard sensors, peripheral switching signals, and timing signals are input to the onboard computer via GPIO. The overall architecture is illustrated in Figure 2.

### *B. Software Architecture*

Based on the above parallel task requirements, the flight software must support multi-task processing and inter-task information exchange. The parallel fault-tolerant flight software integrates primary tasks into independent nodes within the ROS2 framework, achieving information transfer between tasks through the communication mechanisms of topics, services, and actions in ROS2.

It is important to note that, to ensure nodes are started and stopped as required while maintaining control over their lifecycle, all nodes in the flight software are created using the ROS2-specific node type—LifecycleNode. LifecycleNode exhibits multiple states, similar to a finite state machine, and through the ROS2 system function library, it is possible to achieve transitions between the four primary states (Unconfigured, Inactive, Active, and Finalized) and the six transition states, thus meeting the management requirements of the flight software.

The integrated ROS2-based flight software includes the following nodes: timing node, CAN bus task-switching node, payload task nodes, TT&C node, image acquisition node, image processing node, and maintenance node. The functions of these nodes are as follows:

- **Timing Node**: Receives periodic timing signals from the GPS, corrects system time, and publishes periodic time information.
- **CAN Bus Task-Switching Node**: Receives periodic switching signals from the onboard system, manages the switching of payload nodes on the CAN bus, and controls the timing logic of CAN bus access for each payload node.
- **Payload Nodes**: Subscribe to remote control commands from the TT&C node, send information to peripherals via the CAN bus, process data received from payloads over the CAN bus, and publish processed messages.
- **TT&C Node**: Subscribes to processed data from each payload node, stores data, frames telemetry log data for downlink in response to telemetry signals, unpacks remote control commands, and publishes these commands.
- **Image Acquisition Node**: Subscribes to TT&C commands, retrieves images from the CMOS sensor over the MIPI bus under TT&C command control, stores the images, and publishes image data.
- **Image Processing Node**: Subscribes to image data, processes the images, and publishes the processed image information.
- **Maintenance Node**: Subscribes to remote control commands; if maintenance commands are present, this node temporarily takes over message handling for the node requiring maintenance and performs maintenance tasks.

These nodes are interconnected through the ROS2 communication mechanisms, as illustrated in Figure 3.

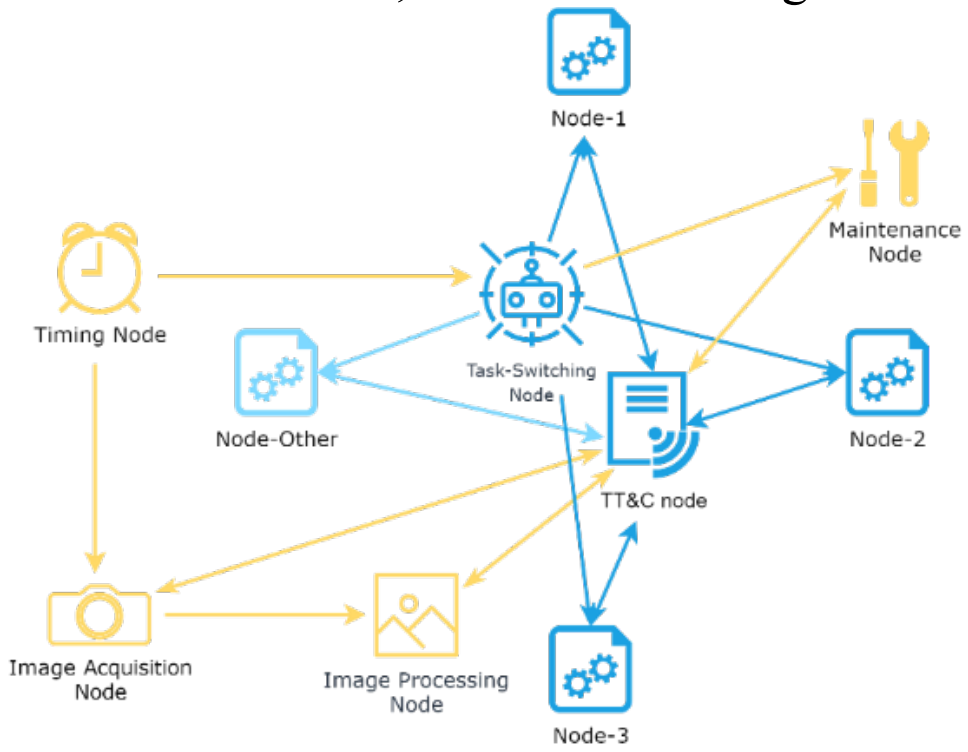

**Fig. 3.** Software architecture of satellite flight software

The arrows shown in Figure 3 illustrate the flow of information between nodes. It is important to note that different communication mechanisms are employed between different nodes: the Timing Node, CAN Bus Task-Switching Node, and TT&C Node utilize the topic mechanism to publish topics, specifically the timing topic, task flag topic, and telecommand topic, respectively. Payload Nodes subscribe to topics published by the CAN Bus Task-Switching Node and TT&C Node using the topic mechanism and publish telemetry topics. Simultaneously, the Payload Nodes communicate with each other using the service mechanism. The Image Processing Node and Image Acquisition Node utilize the ROS2 action mechanism to perform the image acquisition and processing task, functioning as a unified entity for external communication. The Maintenance Node uses parameter communication to interact with the nodes being maintained during node modification tasks.

## IV. Scheduling Process of the Parallel Processing Fault-Tolerance Flight Software

Traditional RTOS and embedded Linux achieve task switching and pseudo-parallelism through interrupts and time-division multiplexing. However, there is currently no comprehensive solution for enabling task interaction within satellite flight software that run multiple parallel programs. Therefore, designing the primary scheduling processes for a parallel processing flight software is essential, including telemetry data collection processes and task uplink and telecommand processes. This scheduling design enables the flight software to effectively perform the three primary tasks described earlier.

### *A. Telemetry Data Acquisition Process*

The telemetry data acquisition process is part of the routine operations of the satellite flight software. The telemetry tasks are initiated in cycles, driven by timing signals from the timing node. Under the control of periodic timing signals from the timing node, the payload switches accordingly.

The switching process of task flag bits is illustrated in Figure 4. The CAN bus task-switching node subscribes to the timing signals of the timing node, and upon receiving each timing signal from the timing node, the switching node initiates a cycle of telemetry data collection for a payload. At this stage, the switching node publishes an array of task flag bit information, with each bit position corresponding to a payload. Each task flag bit has either an active or inactive state. At most, only one bit in the task flag bit array is active at a time, and the active bit sequentially shifts upon receiving a timing signal from the timing node. When the task flag bit corresponding to a payload node is active, the payload is switched to use the CAN bus.

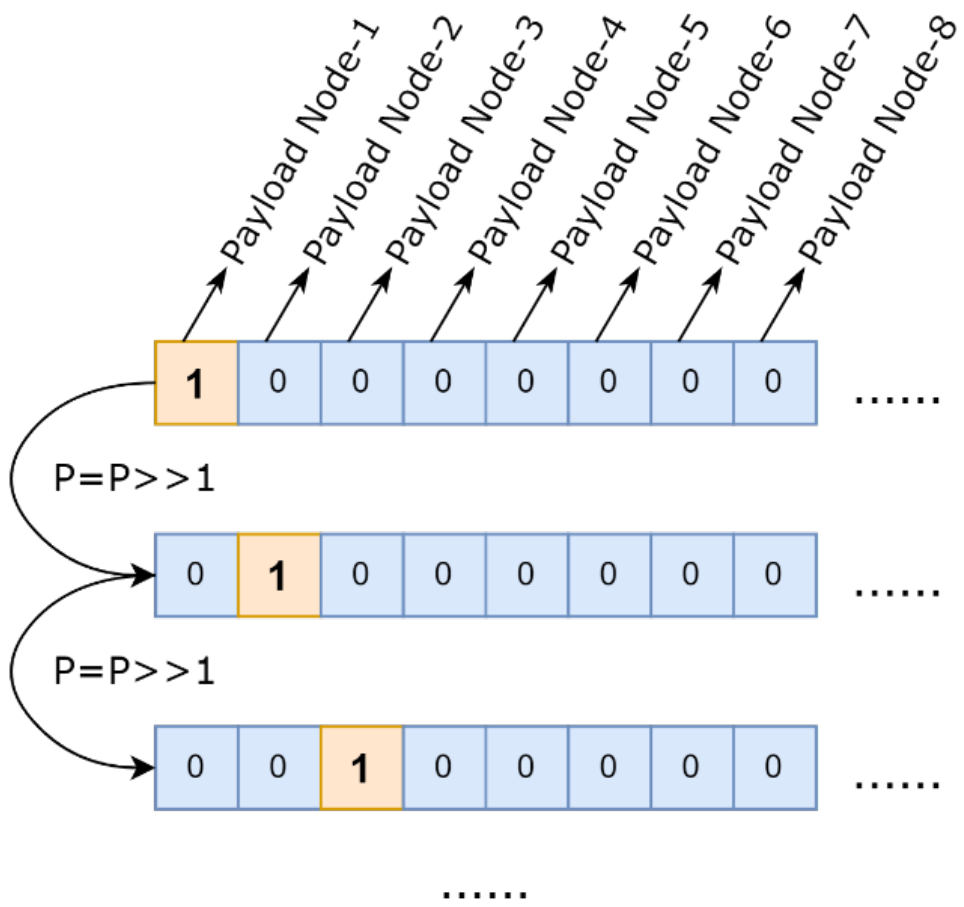

**Fig. 4.** Task flag switching flowchart

The process for subsystem bus access authorization is shown in Figure 5. The payload nodes subscribe to the task flag information from the CAN bus task-switching node. Upon receiving a valid message from this node, the payload node gains permission to use the CAN bus, sending a wake-up signal to the corresponding payload and then waiting for its response data. Once the data from the payload is received, the payload node processes the data and subsequently publishes the processed information.

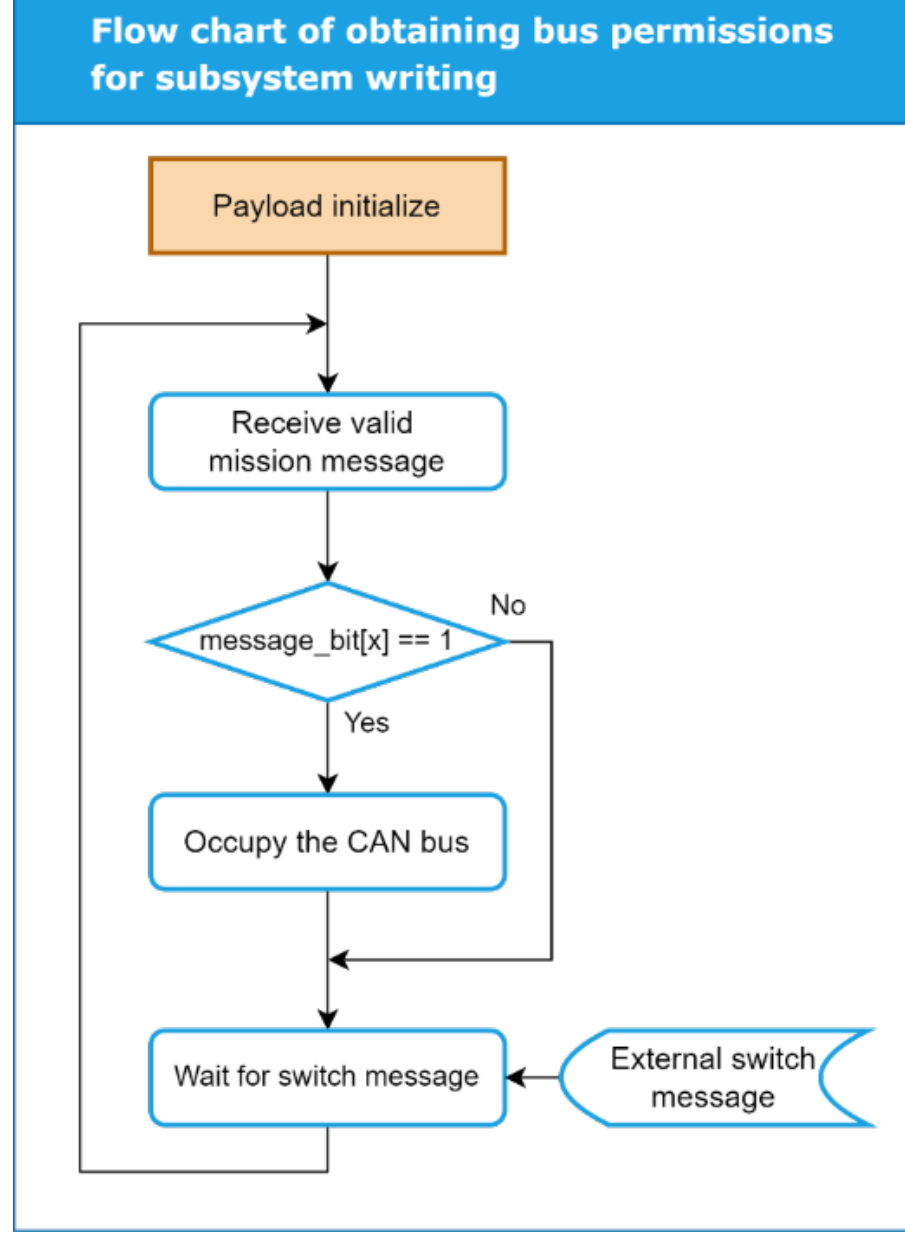

**Fig. 5.** Flow chart of obtaining bus permissions for subsystem writing

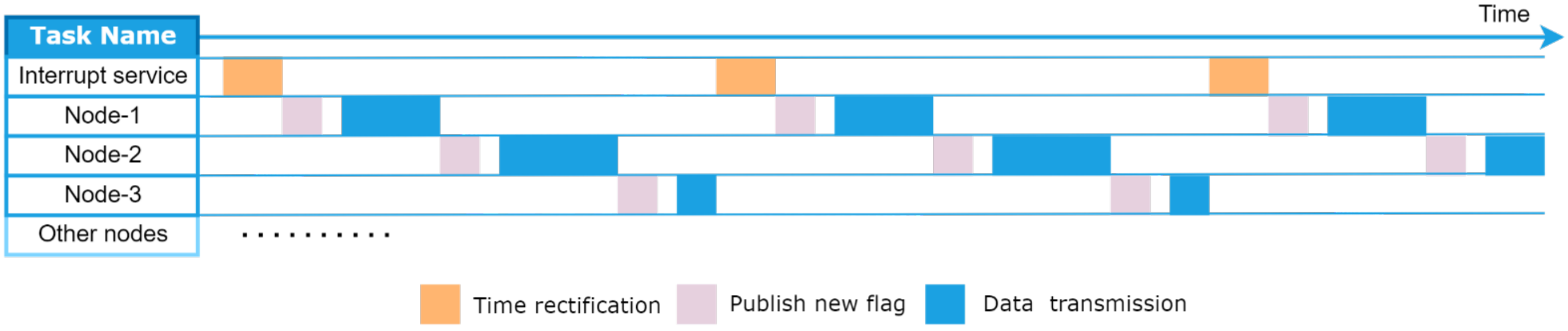

**Fig. 6.** CAN bus time-division multiplexing timing diagram

The TT&C node subscribes to the processed data messages from each payload node. Upon receiving processed data from the various nodes, the TT&C node stores this data and, upon receiving a downlink command from the telemetry and telecommand system, frames the data collected over a specific period and transmits it downward.

Time-division multiplexing of the CAN bus among different payload nodes is a sequential process, as shown in Figure 6. The Timing Node publishes the timing topic at a fixed frequency. When the CAN Bus Task-Switching Node receives the subscription to the timing topic, it checks the usage status of the CAN bus. If the previous payload node has finished its use, the CAN Bus Task-Switching Node will publish the new task flag information. Payload nodes sequentially obtain CAN bus transmission and reception permissions by subscribing to the task flag information in the pre-configured order to acquire payload data. Typically, the pre-set time is sufficient for a node to complete its task. This mechanism is designed based on the watchdog principle, where the CAN Bus Task-Switching Node acts as a central arbiter. If it detects that the task has not been completed within the time allotted by the timing node, it will send a command requesting this peripheral node to restart the corresponding peripheral, stop data transmission, and hand over the CAN bus to the next node for use.

Due to the parallelism of the ROS system, once data acquisition for one payload is complete and data processing begins, the CAN bus can be utilized by another payload node for data transmission and reception. This design improves bus transmission efficiency.

### *B. Task Uplink and Telecommand Process*

The task uplink and telecommand process governs the execution of uplink commands. After unpacking telecommand instructions, the TT&C node publishes the telecommand information. Each payload node subscribes to the telecommand information and, upon receiving the relevant command, initiates the corresponding task to complete the telecommand operation.

As shown in Figure 7, for payloads mounted on the CAN bus, upon receiving a telecommand, the payload node will receive the telecommand information published by the TT&C node, store the command's requirements, and execute the task according to the stored telecommand when it next obtains the CAN bus usage rights.

For image acquisition and processing tasks, when the TT&C node unpacks an imaging task, it first initiates an action client for the image acquisition and processing task. The Image Acquisition Node and Image Processing Node act as action servers. Subsequently, the TT&C node publishes the image task information contained in the telemetry and command. Under the control of the action client, the Image Acquisition Node, upon receiving the subscribed image task information, performs image acquisition at the specified time and publishes the image data. The Image Processing Node, upon receiving the image task information from the TT&C node, selects the image processing method specified by the task and begins subscribing to the image data. After receiving the image data, it starts processing the image and, upon completion, publishes the processed image data. All image acquisition and processing information is aggregated by the image acquisition and processing task action client and sent to the TT&C node. Upon receiving the information, the TT&C node first stores it and, after receiving a downlink command, sends the image data to the telemetry and command subsystem via the RS422 bus.

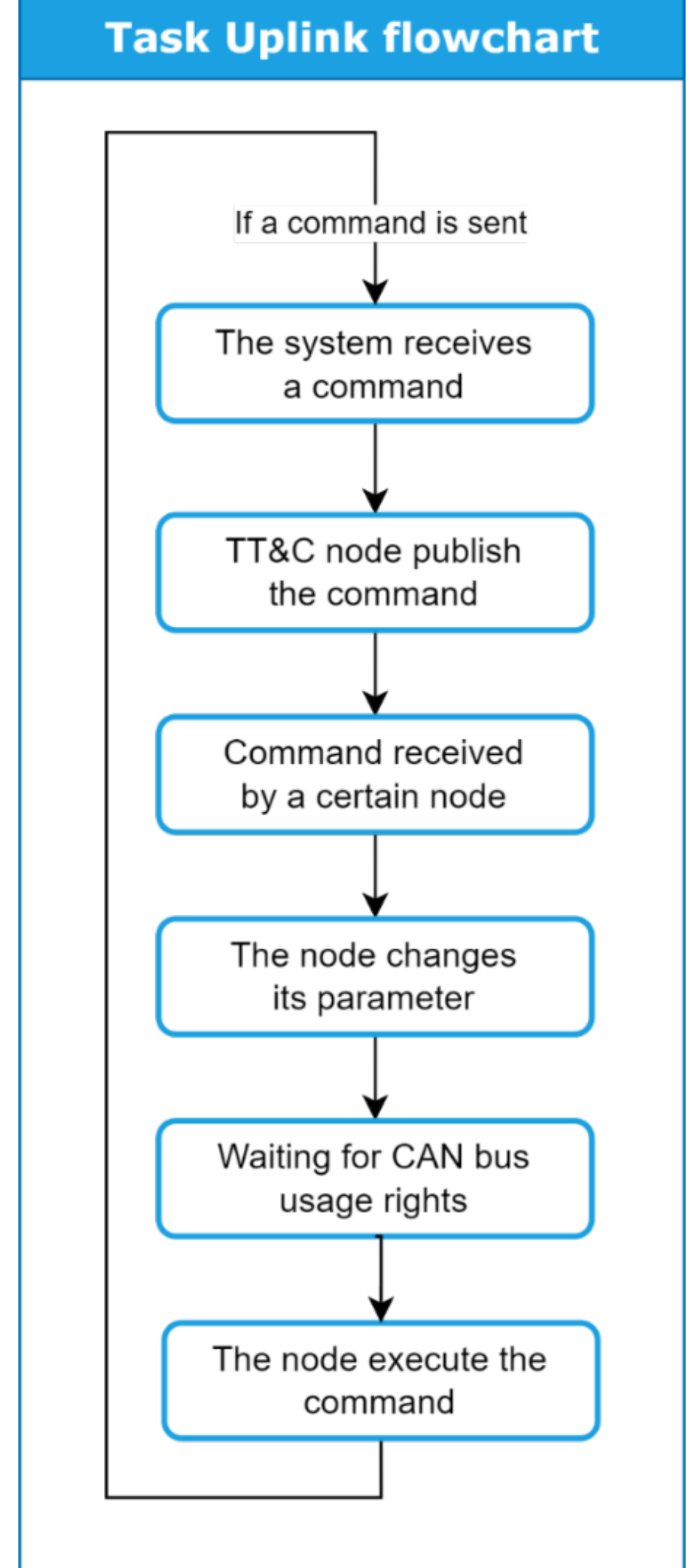

**Fig. 7.** Task annotation flowchart

## V. SIGNAL FLOW LOOP ERROR CORRECTION AND MAINTENANCE MECHANISM

Since the parallel processing fault-tolerant flight software is developed based on the ROS framework, it inherently lacks fault-tolerance mechanisms tailored for satellites. Additionally, microsatellites typically use COTS (Commercial Off-The-Shelf) components, which carry a higher risk of errors compared to space-grade components [32]. Therefore, it is essential to design error correction and maintenance mechanisms specifically for the satellite management system.

The system's error correction and maintenance mechanisms primarily include a heartbeat detection mechanism, a cyclic reboot mechanism, and a node maintenance mechanism. To encompass all nodes in the error correction mechanism, a cyclic reboot mechanism was designed by monitoring the "publish-subscribe" signals, leveraging the distributed nature of ROS nodes. To mitigate the risk of failure in the timing node that drives cyclic error correction, a heartbeat reboot mechanism was also implemented. Furthermore, to meet the system's in-orbit maintenance requirements, a node maintenance mechanism was developed based on signal circulation, enabling the addition and maintenance of nodes. The cyclic reboot mechanism and the node maintenance mechanism address node-level fault tolerance, while the Heartbeat Detection Mechanism provides system-level fault tolerance, thereby implementing a hierarchical fault tolerance design to manage failures at different levels.

### *A. Heartbeat Detection Mechanism*

The heartbeat detection mechanism is implemented through a combination of hardware and software, as shown in Figure 8. After receiving the heartbeat signal from the GPS and completing the timing process, the timing node sends a “feeding” signal to the onboard computer’s watchdog. If the watchdog does not receive the feeding signal within a specified time interval, it will reboot the entire satellite management computer to prevent issues during system operation.

This mechanism integrates the hardware watchdog with the software system, providing a foundational safeguard for the stable operation of the ROS-based system.

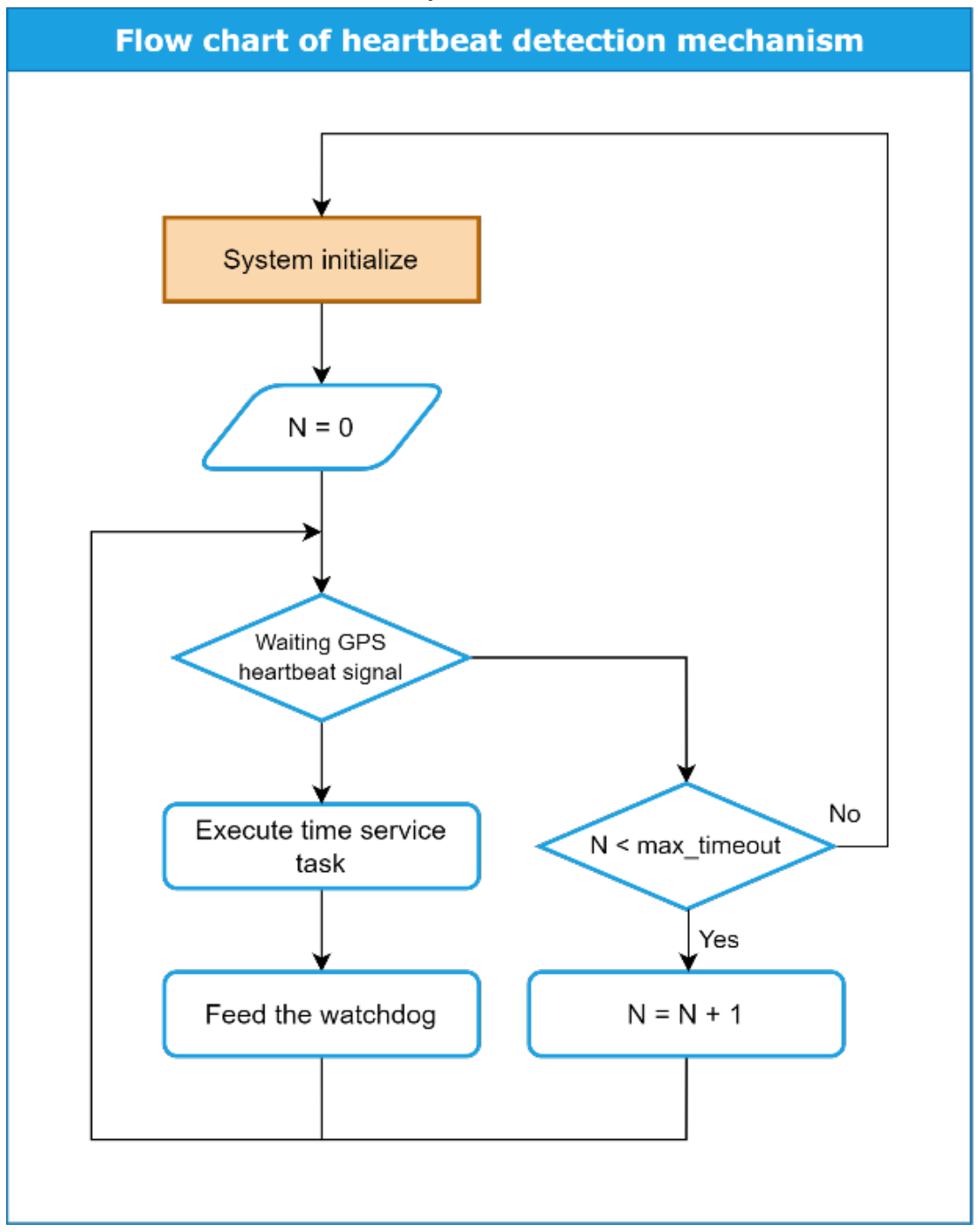

**Fig. 8.** Flow chart of heartbeat detection mechanism

### *B. Cyclic Reboot Mechanism*

In the ROS2 system, software failure manifests as node death. A node may be considered 'dead' if it becomes unresponsive due to various issues. At the software level, this can include an uncaught exception, a segmentation fault, or an infinite loop causing the program to be unresponsive. At the hardware driver level, the driver communicating with a specific payload might freeze due to a temporary fault in a COTS component.

The cyclic reboot mechanism verifies the node's survival status based on the direction of information flow, utilizing ROS2's service communication method. As shown in Figure 9, the upstream node acts as the client in the service, while the downstream node serves as the server. After the upstream node completes its task, it sends a survival status check request to the downstream node. If the downstream node is alive, it will respond to the request; otherwise, it will not. If the upstream node's requests remain unanswered for several cycles, the downstream node is considered dead, and a restart procedure is initiated to restart the downstream node, ensuring the smooth flow of information.

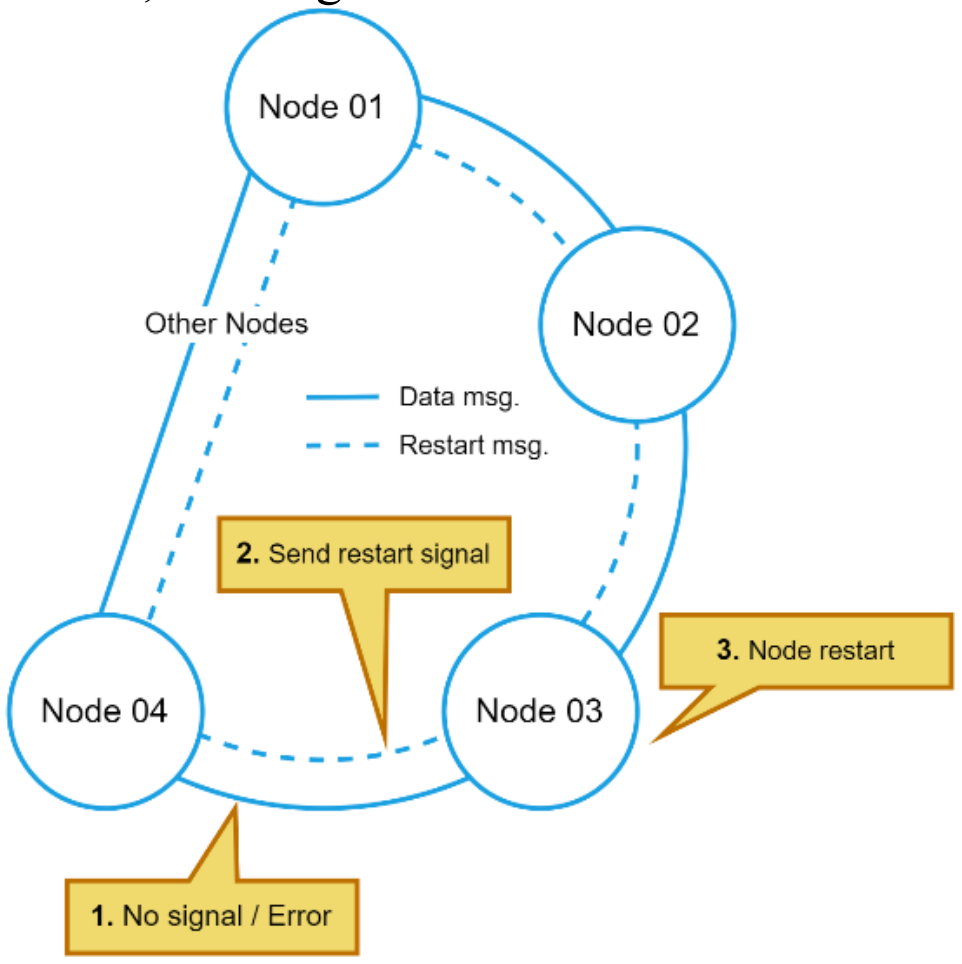

**Fig. 9.** Signal diagram of cycle restart mechanism

This mechanism ensures self-checking and handling of abnormal information flow issues within the system. The chain detection of nodes theoretically guarantees that, even with limited damage, the remaining nodes can still collect data or perform their functions in an orderly manner, regardless of how many nodes the system loses. Moreover, each node can enhance the reliability of its preceding node, making it a symmetric software watchdog implementation.

### *C. Node Maintenance Mechanism*

The node maintenance mechanism is activated by a special telecommand instruction, as shown in Figure 10. When the TT&C node unpacks a node maintenance command, it publishes a maintenance topic to activate the maintenance mechanism. Upon receiving the maintenance topic published by the TT&C node, the maintenance node first determines the type of maintenance command. If it is a parameter maintenance command, the maintenance node modifies the corresponding parameters of the node through parameter communication, ensuring that the node's operation is not disrupted while performing the maintenance. If a new node needs to be added to the system or an existing node needs to be modified, such as adding a new image processing node, the maintenance node will create a new source code file or modify an existing one, then write the code from the command into the file, compile it, and launch the new node. If an existing node is to be modified, the maintenance node will terminate the original node before the new node has been launched, thereby implementing the node modification.

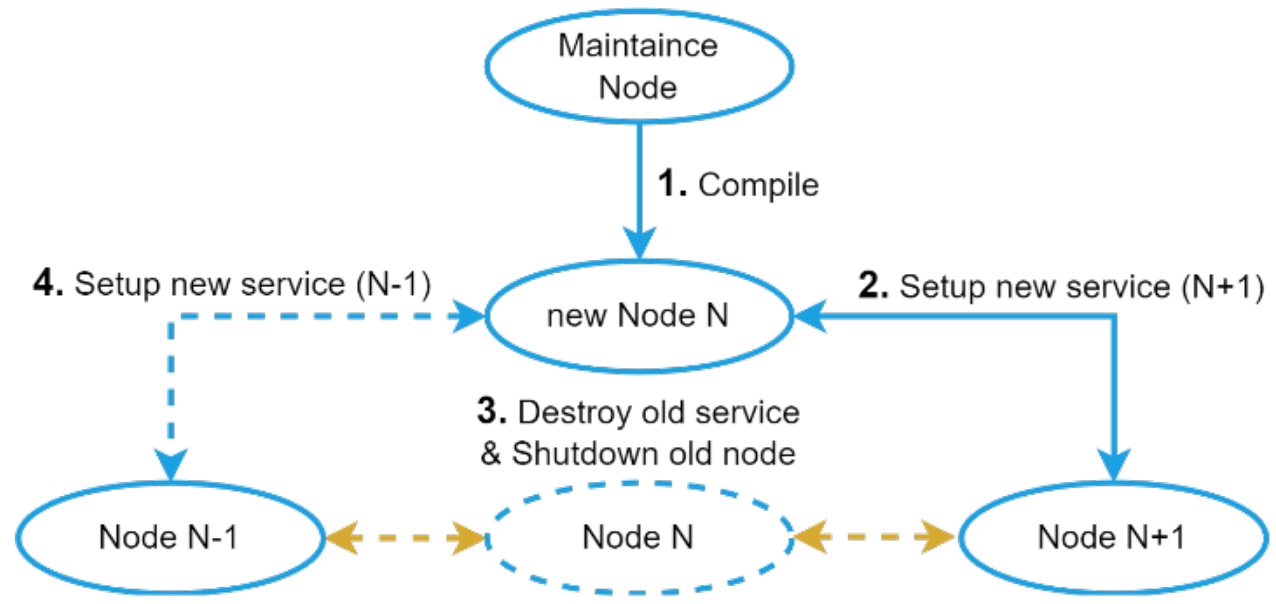

**Fig. 10.** Node maintenance mechanism flowchart.

## VI. EXPERIMENT

Based on the functional requirements of small satellite missions, experiments were designed to assess scheduling latency, system management, fault tolerance, and system maintenance. These experiments enabled the rapid and highly reliable operation of the satellite flight software, validating the system's real-time performance, parallelism, fault tolerance, and maintainability. The test hardware platform utilized the onboard computer of the Space Ranger CubeSat, featuring the RK3568 as the main control chip, as shown in Figure 11. The platform operates on a Linux system with Ubuntu 20.04 as the distribution and uses the Foxy release of ROS2. The three payload nodes (payload01, payload02, payload03) used in the subsequent experiments are virtual nodes designed to emulate the behavior of key satellite subsystems. They are programmed to represent the specific tasks of attitude determination and control, power management, and thermal control systems, respectively.

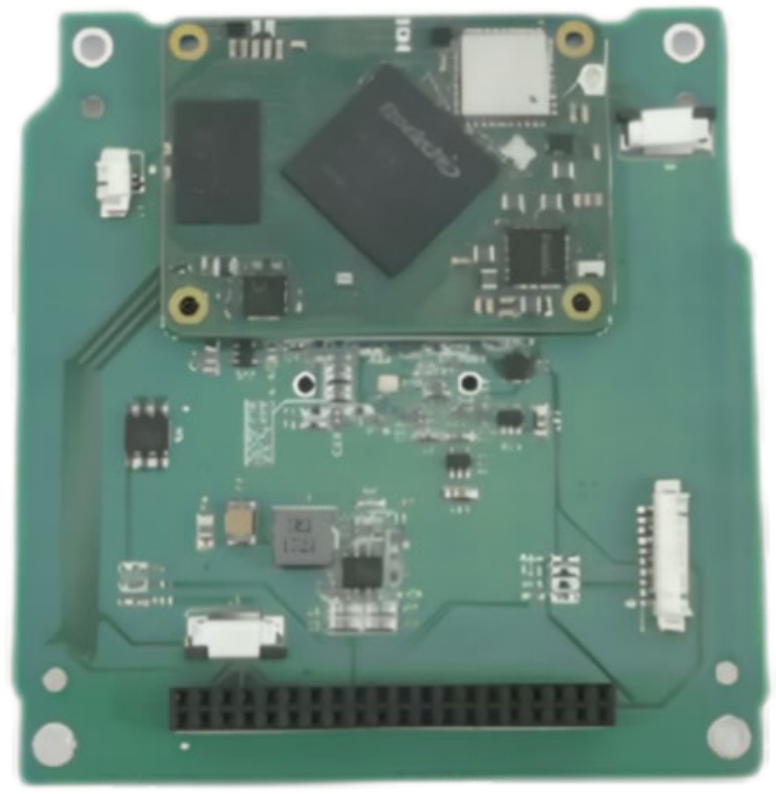

**Fig. 11.** Hardware diagram of testing platform

### *A. Scheduling Latency*

Since the CAN bus task in satellite missions is a periodic task, the software system must meet real-time constraints. Our system is positioned for soft real-time, only needing to meet millisecond-level soft real-time requirements rather than microsecond-level hard real-time requirements; this requirement was found to be sufficient to meet mission demands in actual testing. [33] indicates that a ROS2 environment set up on a Linux system patched with PREEMPT_RT can satisfy real-time requirements. We adopted a similar software stack to build the environment, using Linux kernel version 4.19.232-rt104. The scheduling latency is defined as the temporal deviation between the configured task period and the actual activation time of the task. We evaluated the scheduling latency of the Linux system patched with PREEMPT_RT and the end-to-end latency of the ROS2 publish-subscribe mechanism from publishing to subscription in this environment.

The system's scheduling performance was evaluated using a tool called cyclictest[34]. In our evaluation, we configured a single real-time task to run with a cycle time and deadline of 1 ms, and set the priority to 99 to assess the system's maximum performance. Tests were conducted under both unloaded and fully loaded conditions, with the fully loaded environment simulated using the stress tool stress-ng. In the stressed environment, the CPU continuously runs various computations to simulate heavy computational loads, providing a more accurate reflection of the system's real-time performance under extreme conditions.

Figure 12 presents the results of the system's scheduling latency evaluation. Figure 12a shows the latency results under unloaded conditions, while Figure 12b displays the results under fully loaded conditions. It can be observed that the maximum latency in the unloaded environment is 336 μs, while in the fully loaded environment, it is 94 μs.

This higher latency under light loads is attributed to the processor's power management mechanisms, such as Dynamic Voltage and Frequency Scaling (DVFS) and C-states. Under light loads, the CPU lowers its frequency or enters sleep states to conserve power, introducing wake-up and frequency ramp-up delays when tasks arrive. To verify this, we conducted an additional test by locking the CPU frequency to the maximum value and setting the scaling governor to "performance" for all cores. Figure 12c illustrates the latency under unloaded conditions with these optimizations.

As shown in Figure 12c, the latency distribution significantly improves and aligns with the fully loaded scenario, confirming that the initial delay was due to power management overhead. Although the maximum latency in the default unloaded environment (Fig. 12a) is higher than that in the loaded environment, microsecond-level latency is acceptable in the software system, and the maximum latency does not exceed 500 μs. Therefore, the system can be considered to exhibit real-time performance.

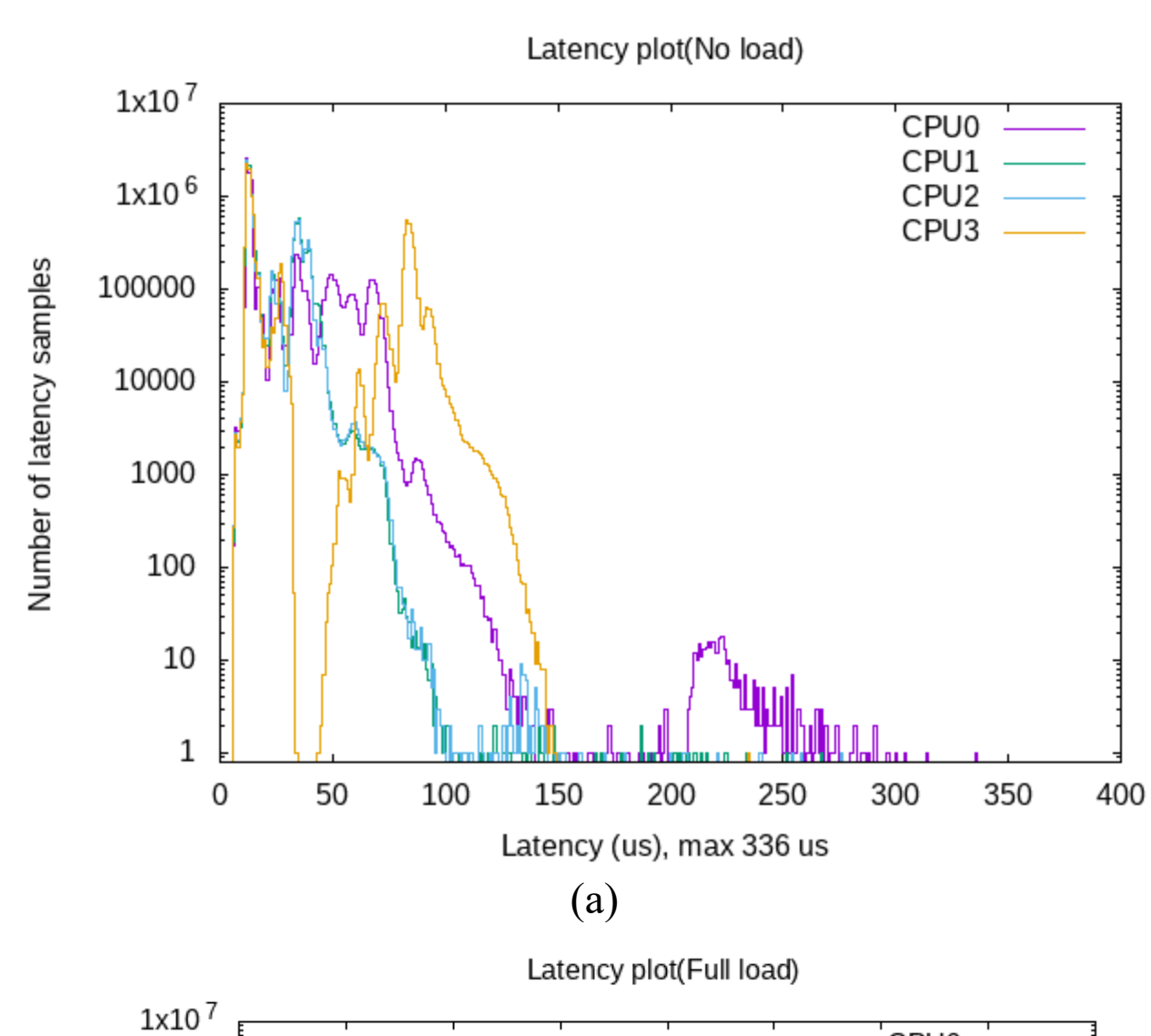

(a)

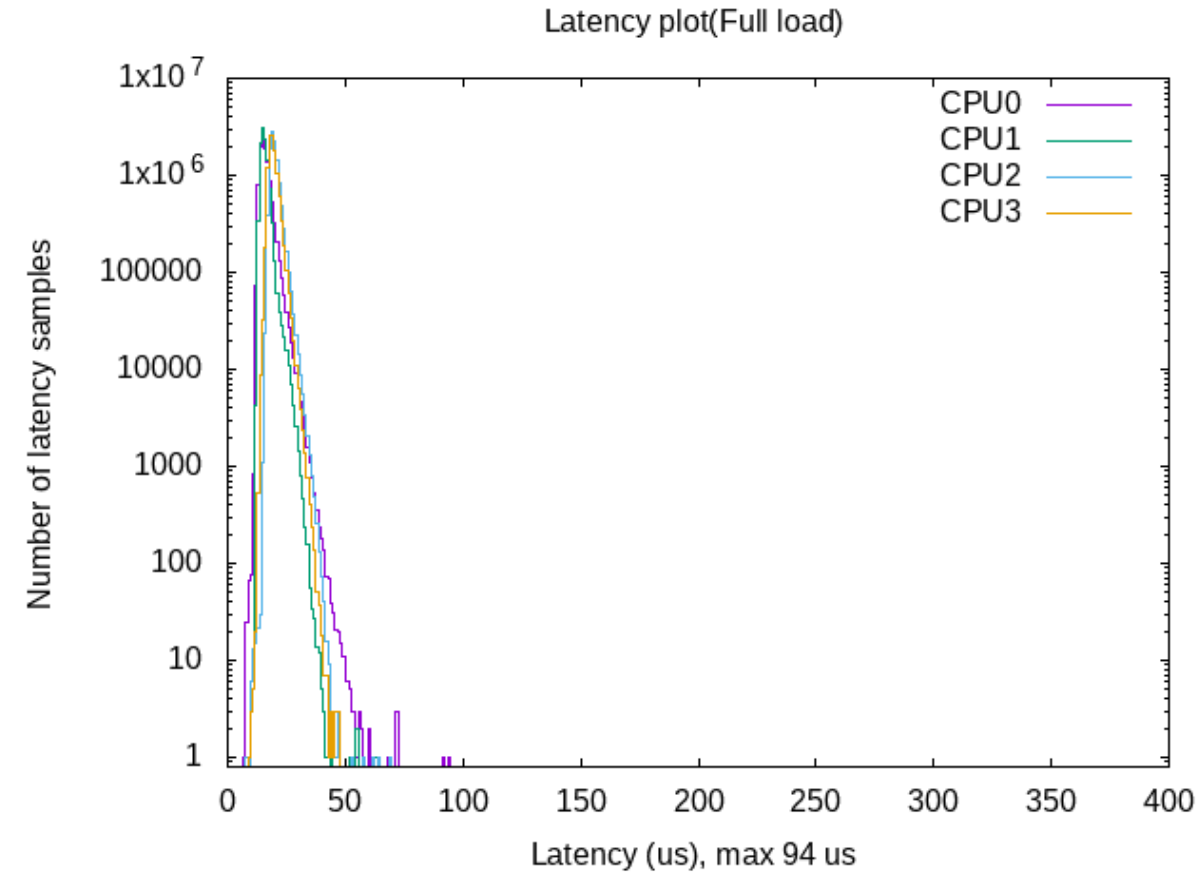

(b)

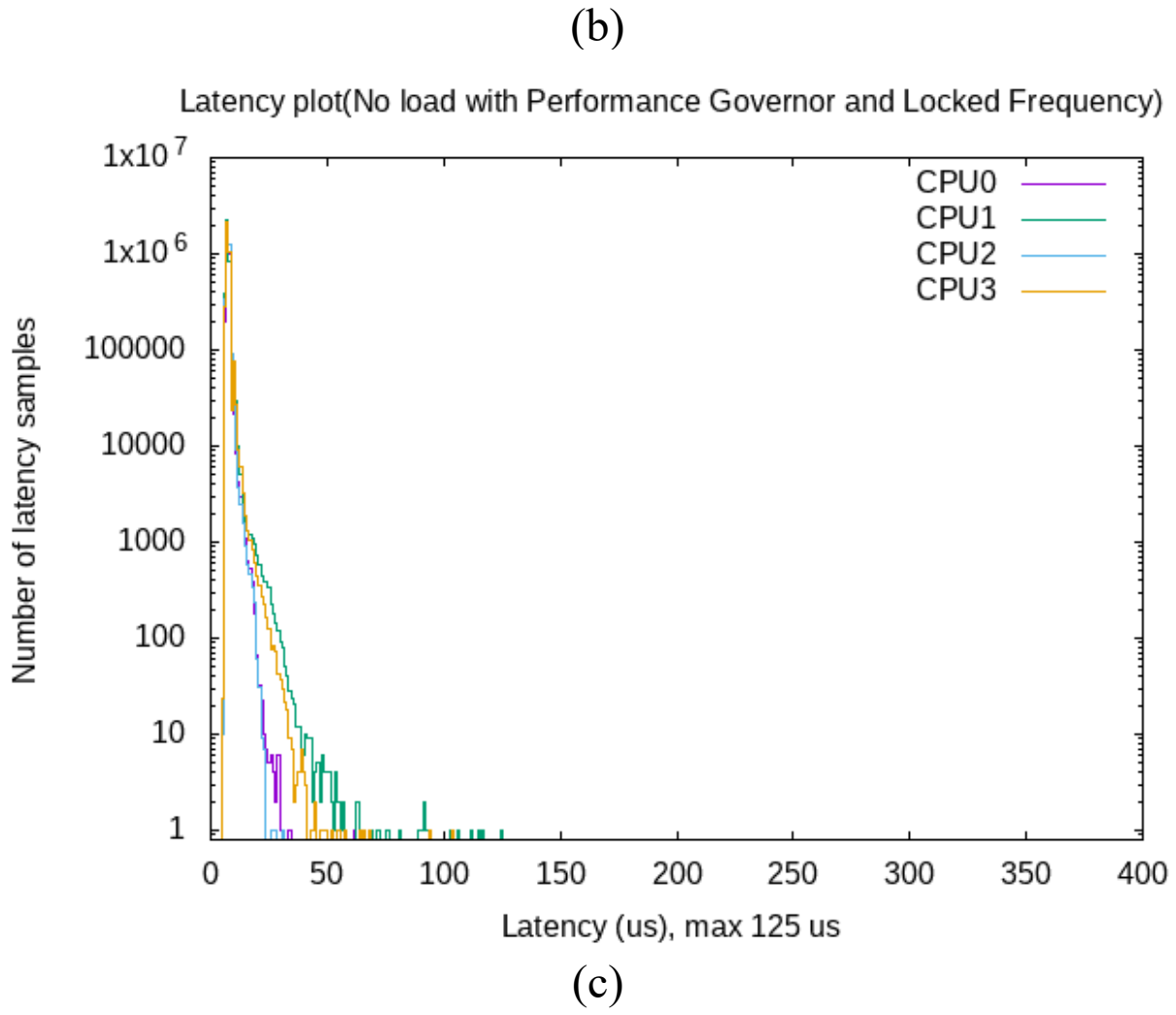

(c)

**Fig. 12.** Scheduling latency of Linux system patched with PREEMPT_RT. (a) No load. (b) Full load. (c) No load with Performance Governor & Locked Frequency

For the latency evaluation of the ROS2 publish-subscribe mechanism, we measured the time delay between message publication and reception by the subscribing node. Since ROS2 logs include built-in timestamps, no additional tools were required for the testing process. Given that only the satellite's mission management component has strict real-time requirements, latency tests were specifically conducted for this component's messages. Two typical scenarios were tested: one with only the mission management component running, representing a light-load state, and the other with both the mission management component and image processing running simultaneously, representing a heavy-load state.

Figure 13 illustrates the test results for the latency of the ROS2 publish-subscribe mechanism. Figure 13a shows the latency results under low-load conditions, while Figure 13b presents the results under high-load conditions. The results indicate that, under both light-load and heavy-load scenarios, the latency fluctuates within a range of a few milliseconds. Under low-load conditions, delays exceeding 10 ms are rare, whereas under high-load conditions, the number of delay occurrences exceeding 10 ms increases. However, the maximum latency in high-load conditions does not exceed 20 ms.

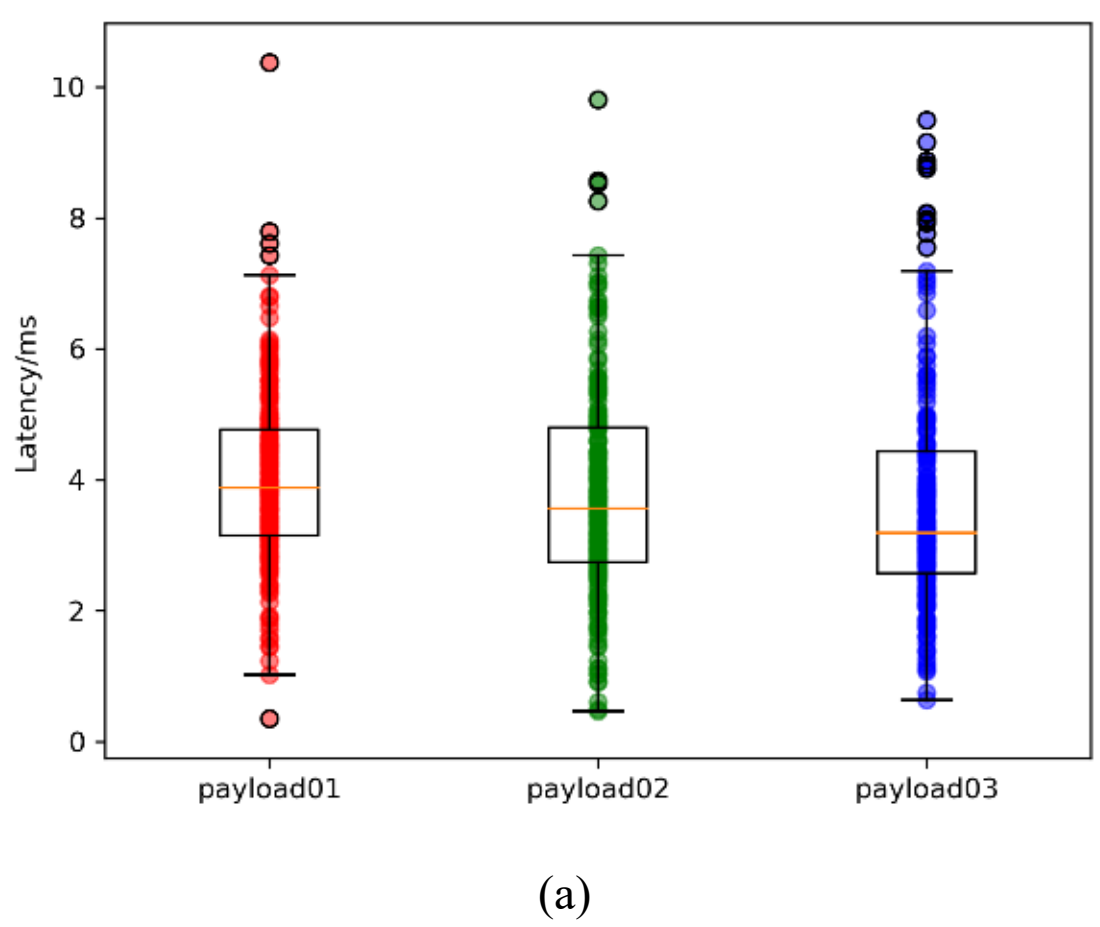

(a)

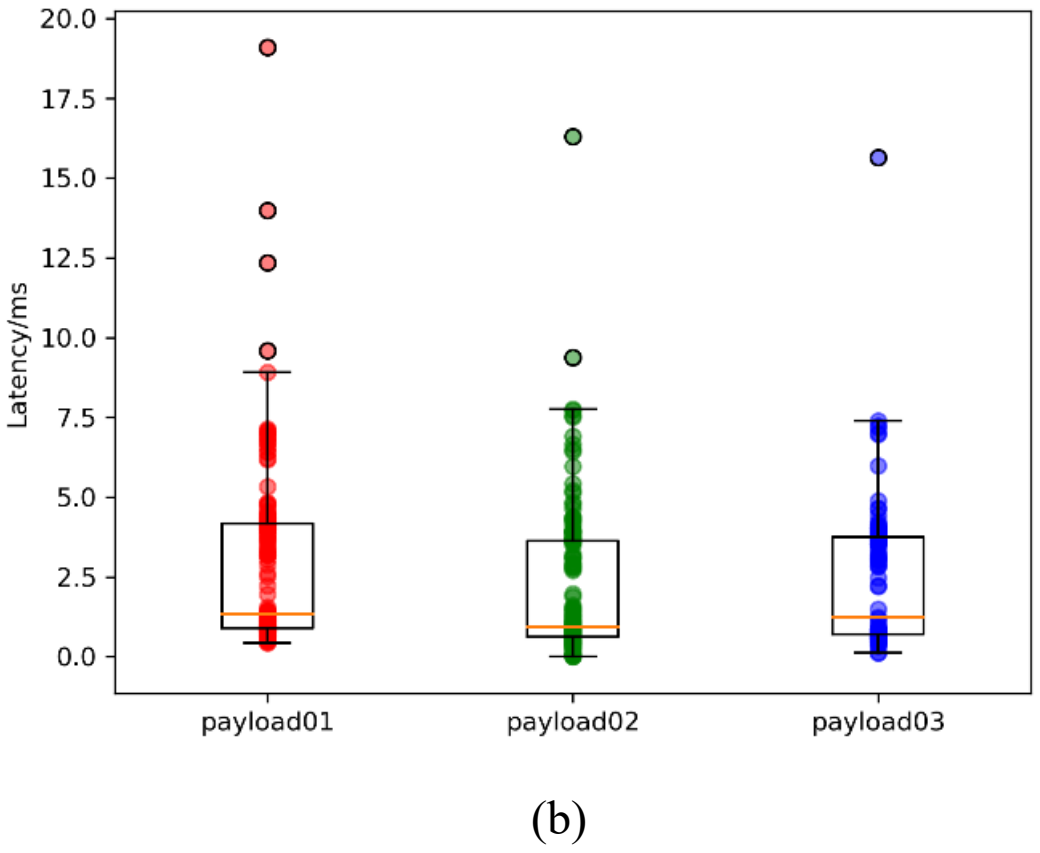

(b)

**Fig. 13.** The latency of the ROS2 publish-subscribe mechanism (a) Light load. (b) Heavy load.

*B. System Management*

The system management experiment primarily validates the effectiveness of the TT&C mechanism under CAN bus switching logic. A simulated satellite management task system was designed, comprising three payload nodes. These payload nodes subscribe to the topics published by the CAN Bus Task-Switching Node and time-share the CAN bus under its control. Upon receiving data from the CAN bus, the payload nodes process the data and publish it to the TT&C node that subscribes to payload information. Additionally, the payload nodes subscribe to telecommand instructions published by the TT&C node.

In our simulation, the payload nodes operate in four states: occupying CAN, data processing, other async commands, and off. Initially, all payload nodes are in the off state. Upon gaining access to the CAN bus, they transition to the occupying CAN state. After finishing their use of the CAN bus, they automatically move to the data processing state for data handling. Furthermore, upon receiving a telemetry command, nodes enter the other async commands state to process the command.

Figure 14 illustrates the different system management states observed in the simulation. Figure 14a shows the normal CAN bus usage state, where the CAN Bus Task-Switching Node switches the CAN bus to the next node at the next heartbeat after each payload node completes its use of the bus. Figure 14b depicts the blocking mechanism during CAN bus switching. In this scenario, payload2 uses the CAN bus for longer than anticipated. At the heartbeat when the CAN Bus Task-Switching Node is supposed to issue a switch command, payload2 is still utilizing the bus. Consequently, the switching of the CAN bus is blocked until payload2 completes its usage, at which point the CAN bus is switched to payload3 in the next cycle. Figure 14c illustrates the parallelism between CAN bus usage and data processing. When payload2 finishes using the CAN bus and transitions to data processing, the CAN bus becomes available for payload3, thereby improving the efficiency of CAN bus usage and task parallelism. Figure 14d demonstrates the handling of telecommand instructions. When the TT&C node issues a telecommand, the corresponding payload node (payload1) briefly transitions to the other async commands state to receive the instruction and executes the command during its next scheduled CAN bus usage cycle.

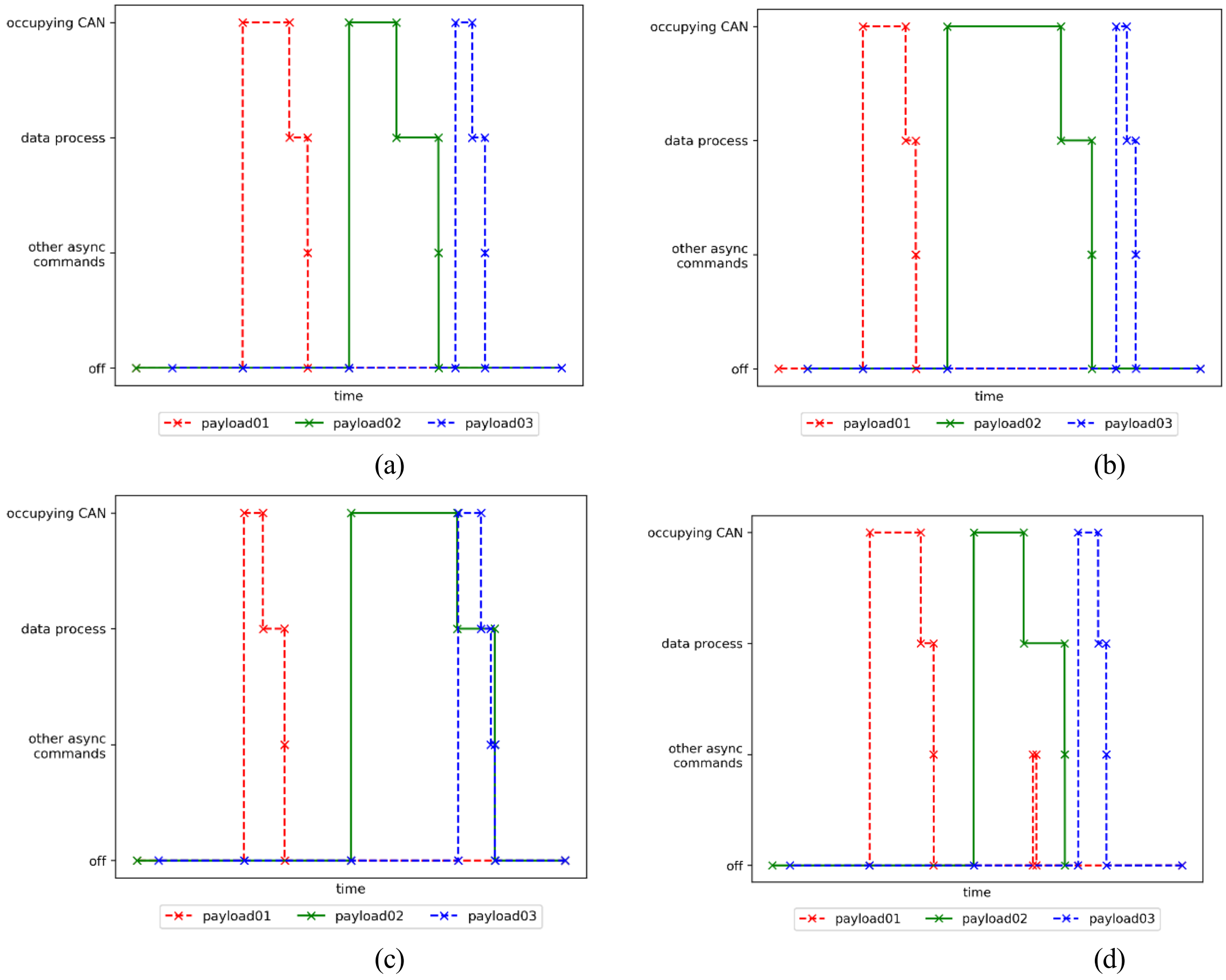

**Fig. 14.** System management states in the simulation. (a) Normal. (b) CAN bus blocked. (c) the parallelism between CAN bus usage and data processing. (d) Receive telecommand. The "other async commands" state is an instantaneous state activated only upon receiving an external asynchronous telecommand. This state is zero in (a)-(c) as they simulate normal, periodic telemetry collection without external command injection, which is the expected behavior. In (d), a telecommand is intentionally injected (on payload1), showing the corresponding brief pulse as the command is processed.

### *C. Fault Tolerance*

The primary fault tolerance mechanism of the flight software is the cyclic restart mechanism. Based on the task simulation system described above, experiments were conducted to validate this mechanism. In the simulation, payload nodes in the other async commands state receive heartbeat signals from downstream nodes. If a node fails to receive a heartbeat signal from a downstream node, it remains in the other async commands state, awaiting the signal. If no signal is received within a specified period, the node increments its no-response counter and shuts down automatically. When the no-response counter reaches three, the node restarts the downstream node to ensure its availability.

Figure 15 presents the results of the simulation experiment. After system initialization, only payload1 is active. Since payload3 is not yet active, payload1 enters the other async commands state but cannot receive a heartbeat signal from payload3. Consequently, after a delay, payload1 exits this state. Once the no-response counter reaches three, payload1 restarts payload3, which then becomes active. After payload3 starts, it fails to receive a heartbeat signal from payload2. Similarly, upon reaching three no-responses, payload3 restarts payload2. This process demonstrates that the entire system can be restarted with only one active node. Additionally, if payload3 unexpectedly fails, payload1 can restart payload3 through the same process, thus ensuring fault tolerance for the failure of a single node.

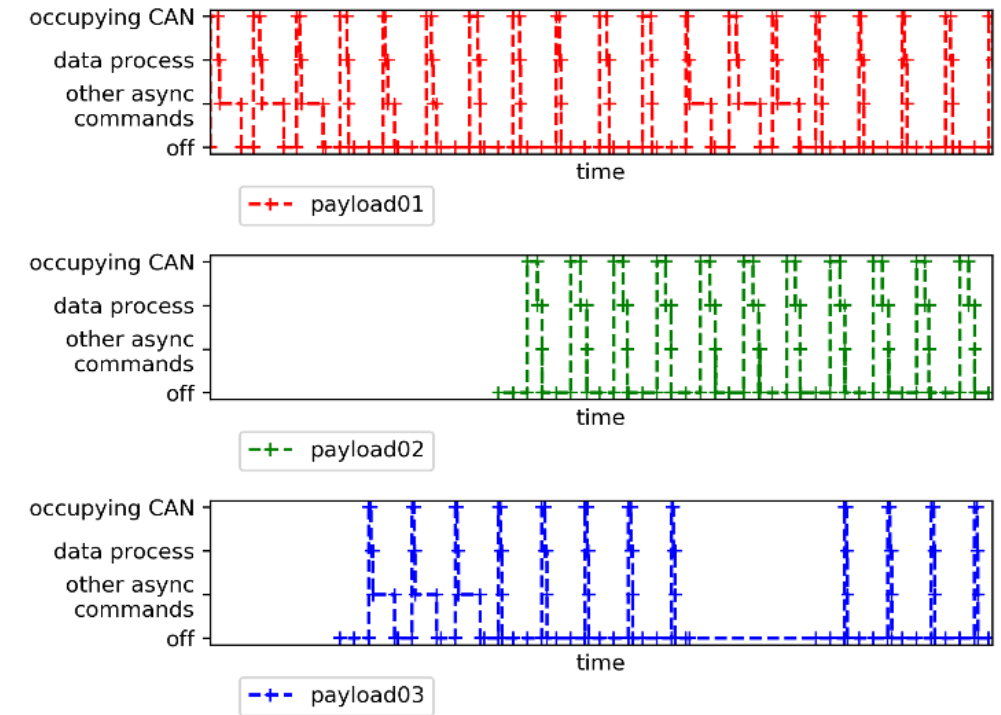

**Fig. 15.** Sequence diagram of node states in the cyclic restart mechanism

### *D. System Maintenance*

For the system maintenance experiment, we designed a scenario where a payload node without CAN bus access is granted permission to use it. This requires the maintenance node to recompile a payload node with the desired functionality and replace the original payload node. Figure 16 illustrates the experimental results: initially, the system had three payload nodes monitoring each other through the Cyclic Reboot Mechanism. The maintenance node deactivated the original payload3, causing payload1 to stop receiving the heartbeat signal from payload3. However, before payload1 triggered the mechanism to reboot payload3, the maintenance node successfully compiled and launched a new payload3, thereby completing the node maintenance process.

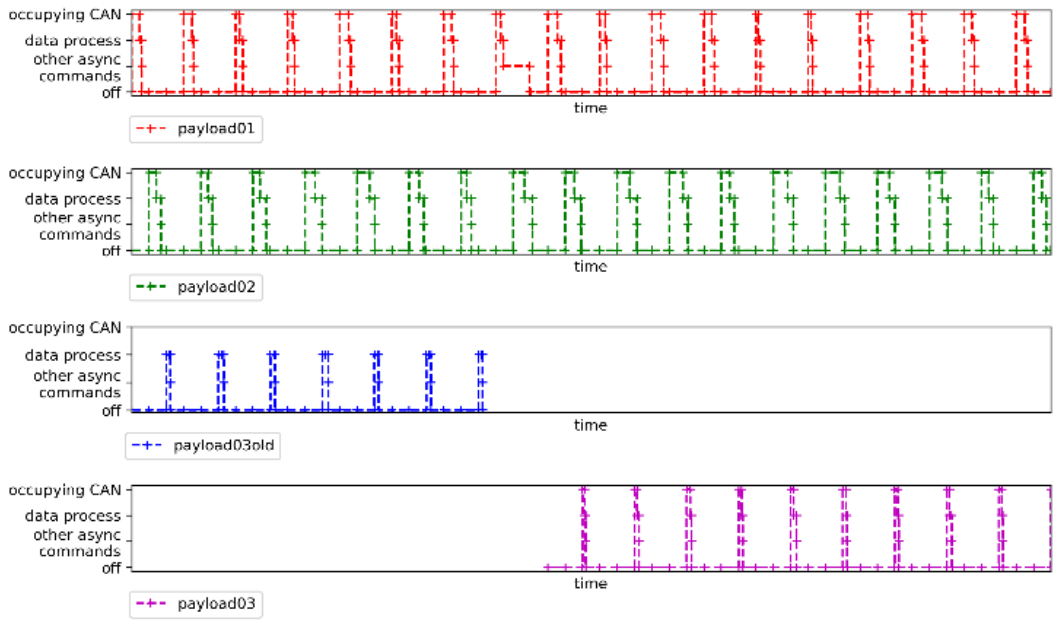

**Fig. 16.** Sequence diagram of node states in the system maintenance mechanism

## VII. Discussion

Regarding system time delays, the current average delay is a few milliseconds, with a maximum of no more than 20 milliseconds. This is acceptable for our satellite software system; however, spacecraft systems requiring higher time precision, such as inertial navigation systems, may necessitate low-level modifications to the ROS2 system to meet stringent delay requirements. [36] suggests that redesigning the ROS2 executor to optimize scheduling mechanisms can achieve sub-millisecond delays, which may represent a future direction for spacecraft software systems based on ROS2.

Additionally, an interesting phenomenon was observed during the time delay experiments: the system's real-time performance was better under high CPU load conditions than under low CPU load conditions. Our additional experiment (Figure 12c) confirmed that this is related to the processor's power management state. Under low load, CPU cores may frequently enter deep sleep states. When a task needs to run, waking the core from deep sleep requires extra time, which introduces latency. Under high load, CPU cores remain in an active state, avoiding wake-up latency. By locking the frequency and enabling the performance mode, we demonstrated that this latency can be eliminated, though this comes at the cost of higher power consumption. Consequently, the average latency for task scheduling and response may actually be lower. This is consistent with analyses from the Realtime Linux community[37].

Regarding system communication, in the design of the flight software, we extensively adopted ROS2's new communication mechanisms to accomplish complex tasks and logic with less effort. Compared to traditional satellite software systems like cFS and F prime, ROS2, as an open-source robotics system, benefits from a global community of researchers actively using and enhancing it. This results in faster software iteration and quicker implementation of new features. Additionally, ROS boasts a large community and extensive open-source resources, providing valuable references for developing specific satellite missions and facilitating rapid iteration in the development of small satellites. From a software engineering perspective, ROS is relatively easy to learn and develop with, while its distributed architecture simplifies the maintenance of various system components.

Regarding system maintenance, ROS2's loosely coupled mechanism based on DDS allows for decoupling between different parts of the system, making it possible to independently modify, compile, and run individual components, which is not achievable in other software systems.

## VIII. Conclusion

This paper presents a parallel and fault-tolerant ROS2-based flight software for small satellites, addressing the issues of poor parallel performance in traditional real-time operating systems and the difficulty of thread monitoring in Linux-based systems. Through the design of the hardware-software architecture, parallel system scheduling mechanisms, and signal flow cyclic error correction and maintenance mechanisms, this system meets the parallel and fault-tolerant task requirements of small satellites. Verification on a ground test platform demonstrates that the parallel fault-tolerant satellite operations system has significant advantages in terms of parallel processing capabilities, system fault tolerance, and system maintenance complexity.

Compared to existing satellite flight software, this approach only requires consideration of the information flow logic during system development, reducing the emphasis on the timing and priority of information transmission. This represents an innovative approach in the development of satellite operations systems. In the future, further exploration will be conducted on the application of ROS2-based satellite operations systems on different types of satellites, with optimizations aimed at improving system performance under varying task requirements.

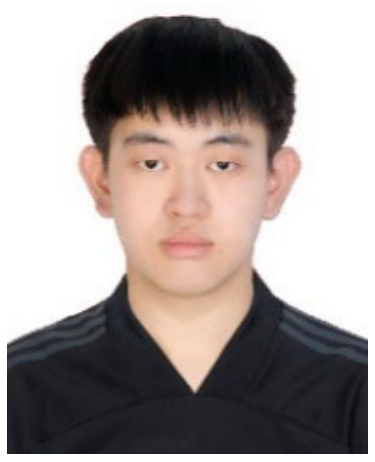

**Zebei Zhao** was born in Nei Mongol, China, in 2003. He received the B.S. degree in Automation from Beihang University, Beijing, China, in 2025. He is currently pursuing the M.S. degree in Electronic Information with the School of Artificial Intelligence and Data Science, University of Science and Technology of China, Hefei, China. He previously served as the group leader of the Flight Software Group within the Space Ranger Satellite High-Level Student Science and Technology Innovation Team at Beihang University. His research interests include flight software, large language model and reinforcement learning.

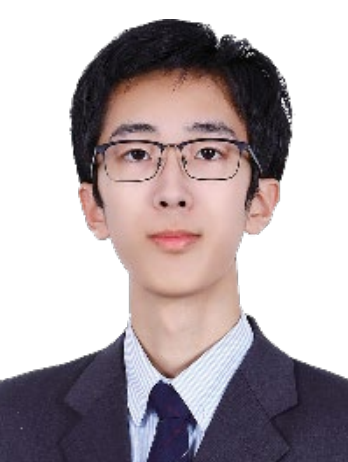

**Yinghao Xiang** was born in Sichuan Province, China, in 2004. He is currently an undergraduate majoring in Computer Science and Technology at Beihang University, Beijing, China. He previously served as the deputy group leader of the Flight Software Group within the Space Ranger Satellite High-Level Student Science and Technology Innovation Team at Beihang University. His research interests include flight software, deep learning and computer vision.

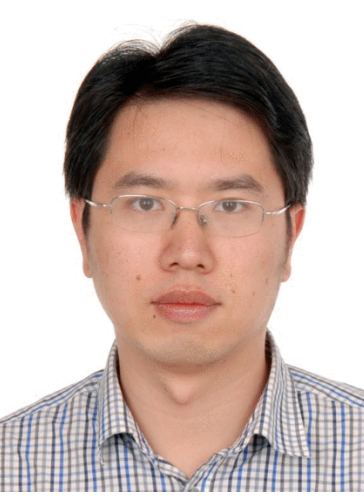

**Pei Chen** was born in Beijing, China, in 1979. He received the Ph.D. degree in aerospace engineering from Beihang University, Beijing, China, in 2008. He is currently an Associate Professor with the School of Astronautics, Beihang University. His current research interests include spacecraft navigation, GNSS application, and astrodynamics and simulation.